# A CONJECTURE CONCERNING THE RIEGER AND QUASI-BIENNIAL SOLAR PERIODICITIES

P.A. Sturrock
Center for Space Science and Astrophysics,
Stanford University, Stanford, CA 94305

## ABSTRACT

It has been established that the Rieger periodicity of approximately 153 days is part of a complex of periodicities, all multiples of a basic period of approximately 25.5 days. However, it has not been clear why the sixth subharmonic of this periodicity should be preferentially manifested. We here note that if the Sun contains two rotating elements, with different periods and different axes, a special situation will arise if the two periods have a lowest common multiple, for in this case the relative configuration of the two rotators would repeat exactly at that (LCM) period. This chain of thought leads us to suspect that the Sun contains a second rotating element with rotational period in the range 21 - 22.5 days.

## 1. INTRODUCTION

Rieger et al. (1984) presented evidence that gamma-ray flares exhibit a periodicity of order 153 days, which we here refer to as the "Rieger" periodicity. Subsequent study has shown that this periodicity is present also in other forms of solar activity (see, for instance, Bai & Cliver 1991, Kile & Cliver 1991). Bai and Sturrock (1993) have shown that the Rieger periodicity is part of a complex of periodicities that are approximate multiples of a "fundamental" period of 25.5 days, and also that analysis of the timing and location of major flares provides supporting evidence for the proposition that the Sun contains an oblique rotator with this period. The term "oblique rotator" here signifies a region that rotates about an axis that differs from the axis of the Sun's surface rotation.

However, it is curious that the Rieger periodicity is the dominant member of this complex of periodicities: it is a puzzle to understand why the sixth subharmonic should be preferentially manifested. The purpose of this Letter is to suggest an answer to this question, and then to examine some of the consequences of this suggestion.

## 2. PROPERTIES OF A TWO-ROTATOR SYSTEM

If we are to understand the Rieger complex of periodicities, it seems to be necessary to extend the earlier model of a single "fundamental" periodicity (that may or may not be due to a rotating element). We therefore consider the next simplest possibility, namely that some part of the Sun contains *two* interacting dynamical components. If each of these has an intrinsic periodicity, then the interaction could result in a rich array of periodicities and quasi-periodicities. Each component could in principle be either a rotating element or an oscillating element. An example of the latter would be a torsional oscillator of the type proposed by Walen (1947) in his model of a possible mechanism of the solar cycle.

While recognizing that each component may be either a rotator or an oscillator or something more complicated, we here focus on the specific possibility that the two components are two interacting regions of the solar interior, that we refer to as "rotators", that have different rotation periods, and that we allow also to have different rotational axes. A specific possibility is that the inner rotator, which contains most or all of the nuclear burning region, rotates with period $P_1$ about one axis, and the outer rotator, which may comprise an outer part of the core and/or an inner part of the radiative zone, rotates with period $P_2$ about a different axis. Each rotator may be a substantial part of the solar interior - most of the core, or most of the radiative zone - or perhaps only a small part of the interior, just as the atmospheric Jet Stream or the oceanographic Gulf Stream is only a small part of the structure of the Earth.

Since the inner rotator loses angular momentum only by transfer to the outer rotator, we must require that $P_1 < P_2$. We suppose that both rotators are inhomogeneous due to variation in composition, due to magnetic field, or for some other reason, and we suppose that the inhomogeneities are sufficiently large that the interaction of the two rotators leads to variation in the nuclear burning rate. In this scenario, hydrodynamic (a term that here includes magnetohydrodynamic) motions are modulating the energy generation rate. Conversely, time-varying and asymmetric energy generation leads to inhomogeneity of the core and to hydrodynamic flows. The term "engine," that has been used in referring to the solar interior (see, for instance, Nesme-Ribes 1993), may truly be appropriate.

Clearly, the first step in developing this line of thought is to find evidence that there are two (or more) "fundamental periodicities" involved in the solar interior - meaning the region inside the convection zone. This requires the specification of a "search band" for the second periodicity. If there is no evidence for a second periodicity, there is no need to explore these concepts further. If evidence for a second periodicity is forthcoming, then there will be good reason to further explore these concepts. If we adopt as one of these periods that which has already been identified, in the range 25.4 - 25.6 days, we are left with the challenge of finding the second periodicity. In this Letter, we seek to narrow the range of possible values of the second period.

One must expect that the output from a two-rotator sytem would vary with time in a complex manner. There may be some evidence of both periodicities $P_1$ and $P_2$, but in general the system will have no overall strict periodicity. An interesting situation arises, however, if the two periods are commensurable or approximately commensurable. Suppose, for instance, that $P_1$ and $P_2$ have an exact lowest common multiple $P_L$. Then after a time interval $P_L$, the relative configuration of the two rotators is returned *exactly* to its state at the beginning of the time interval: in this case there will be an exact periodicity with period $P_L$.

These considerations lead to the following conjecture concerning the Rieger periodicity. We suppose that the two periods are approximately commensurable so that $P_R$ is, to good approximation, an integral multiple of both $P_1$ and $P_2$:

$$P_R \approx n_{R1} P_1, \quad P_R \approx n_{R2} P_2 . \tag{1}$$

One of these periods should be close to the value (25.5 days) proposed by Bai and Sturrock (1993), and this points to the choice $n_{R2} = 6$. We now need to explore the possible values of $n_{R1}$.

We can consider in turn the possibilities $n_{R1} = 7,8,9,10,11,...$ Of these, the choices $n_{R1} = 8, 9$ or $10$ would not make sense, since the common value of $n_{R1}P_1$ and $6P_2$ would then not be the lowest common multiple of $P_1$ and $P_2$. If we must choose between $n_{R1} = 7$ and $n_{R1} = 11, 13$, etc., it is clearly more conservative to adopt the former, since this leads to the lowest acceptable rotation rate for rotator 1 in the context of this model. We therefore examine the possibility that $n_{R1} = 7$: this leads to the hypothesis that the inner rotator has a period of order 153/7 days.

In order to arrive at a more specific "search band" for $P_1$, we have examined the time series provided by the sunspot number for the time interval 1850 to 1970, which provides a much longer usable record than does, say, the record of solar flares. We find that the spectrum of this time series exhibits a prominent peak at 152.3 days, with "wings" extending up to 157 days and down at least to 147 days. By adopting $P_R \approx 152 +/- 5$ days, we obtain the estimate $P_1 = 21.7 +/- 0.7$ days. The choice $n_{R2} = 6$ leads to $P_2 = 25.3 +/- 0.8$ days, which brackets both the value of 25.5 days proposed by Bai and Sturrock (1993) and also the nearby (sidereal) Carrington period (Bruzek & Durrant 1977) of 25.38 days. (It is well known that the synodic value of the Carrington period [27.275 days] figures prominently in the time variation of solar activity. See, for instance, Kundt 1993.) These considerations lead one to consider the possibility that the inner rotator has a period in the range 21.0 days to 22.4 days, and that the outer rotator has a period in the range 24.5 days to 26.1 days.

## 3. DISCUSSION

One test of this model is to search for evidence (including harmonics and sub-harmonics) of a periodicity in the range proposed for $P_1$. It is notable that a recent analysis of solar-wind data by Thomson, McPherron and Lanzerotti (1995) revealed a very prominent peak at a period of about 10.6 days which, as pointed out by those authors, occurs also in other solar-activity data. These signals may prove to be the harmonic of $P_1$, the hypothesized rotation period of the core. If this proves to be the case, we will find that $P_1$ is close to 21.2 days. It will be interesting to learn whether further analyses of solar wind and related data sets yield a periodicity near 21 days in addition to the periodicity near 10.6 days. We note that the periodicity $P_1$ may also show up indirectly through combination frequencies such as $\nu_1 - \nu_2$, $\nu_1 + \nu_2$, etc.

Current information concerning the internal rotation of the Sun is somewhat confusing. On the one hand, there are claims that the period of rotation of the core may be quite short (Toutain and Froelich 1992), and on the other hand there are claims that it may be quite long (Elsworth et al. 1995 ). Clearly, the present proposal will ultimately stand or fall on more definitive measurements of the internal rotation rates. The recent commissioning of the Global Oscillation Network (Harvey et al. 1996), and the recent launch of the Solar and Heliospheric Observatory that carries several helioseismology experiments (Frohlich et al. 1996, Gabriel et al. 1996, Scherrer et al. 1996), will lead to helioseismological data of unprecedented duration, accuracy and stability that will make it possible to probe much more deeply into the solar interior. These measurements will hopefully yield definitive estimates of the structure and internal rotational velocity of the solar core.

Another test of this model is to search for evidence that the Rieger periodicity originates deep in the Sun, rather than near the surface. It is in this context notable that, according to Ribes et al. (1989), the spectrum of diameter measurements exhibits the Rieger periodicity, a result that seems easier to reconcile with a deep-seated origin of the periodicity than with a near-surface origin.

The Rieger periodicity is not the only enigmatic solar periodicity. The solar cycle, with a total period of 22 years, is generally attributed to a dynamo process in the solar convection zone (Krause et al. 1993; Proctor et al. 1995), although this theory is not without its difficulties (Kundt 1993; Schmitt 1993; Sokoloff, D., et al. 1993).

Another intriguing cycle is the quasi-biennial periodicity (see, for instance, McIntyre 1993, Ribes 1989) with a period $P_Q$ of approximately 26 months. If evidence for $P_1$ is forthcoming, it will be interesting to examine the possibility that the quasi-biennial periodicity also may be interpreted in a way similar to that proposed for the Rieger periodicity. That is, we should examine the possibility that $P_Q$ is integrally (or approximately integrally) related to both $P_1$ and $P_2$:

$$P_Q \approx n_{Q1} P_1, \quad P_Q \approx n_{Q2} P_2 . \tag{2}$$

In this context, it is interesting to recall that Sakurai (1979, 1981) has claimed to find evidence that the solar neutrino flux exhibits the quasi-biennial periodicity. This claim, if true, could be reconciled with a process originating in the solar core more readily than with a process confined to the convection zone.

The current consensus (see, for instance, Bahcall 1989, Bahcall et al. 1996), is that variation of the measured neutrino flux may be attributed to the effect of the solar magnetic field on the propagation of neutrinos that have non-zero mass and non-zero magnetic moment. However, we should perhaps be cautious and not neglect the possibility that nuclear burning may be inhomogeneous and time-varying, as recently suggested by Grandpierre (1996).

Evidence for a periodicity of order 21 days, in data relevant to processes in the solar core, is presented in the accompanying Letter (Sturrock & Walther 1996).

I thank many friends for helpful discussion, including Taeil Bai, Douglas Gough, George Roumeliotis, and Michael Wheatland. This work was supported in part by Air Force grant F49620-95-1-008 and NASA grant NAGW-2265.

## REFERENCES

Bahcall, J.H. 1989, Neutrino Astrophysics (Cambridge Univ. Press).

Bahcall, J.N., Calaprice, F., McDonald, A.B., & Totsuka, Y. 1996, Physics Today (July 1996), 30.
Bai, T., & Cliver, E.W. 1991, ApJ, 363, 299.
Bai, T., & Sturrock, P.A. 1993, ApJ, 409, 476.
Bruzek, A., & Durrant, C.J. 1977, Illustrated Glossary for Solar and Solar-Terrestrial Physics (Dordrecht, Holland: Reidel), 11.
Elsworth, Y., et al. 1995, Nature, 376, 669.
Frohlich, K., et al. 1995, Solar Phys., 162, 101.
Gabriel, A.H., et al. 1995, Solar Phys., 162, 61.
Grandpierre, A. 1996, Astron. Astrophys. 308, 199.
Krause, F., Raedler, K.-H., and Ruediger, G. 1993, The Cosmic Dynamo. Proc. 157th IAU Symp, Potsdam, September 7 - 11, 1992 (Dordrecht: Kluwer).
Harvey, J.W., et al. 1996, Science, 272, 1284.
Kile, J.N., & Cliver, E.W. 1991, ApJ, 370, 442.
Kundt,W. 1993,The Cosmic Dynamo. Proc. 157th IAU Symp, Potsdam, September 7 - 11, 1992 (eds. Krause, F., Raedler, K.-H., and Ruediger, G.; Dordrecht: Kluwer), 77 .
McIntyre, M. 1993, The solar engine and its influence on terrestrial atmosphere and climate (ed. E. Nesme-Ribes; Berlin: Springer), 293.
Nesme-Ribes, E. (ed.) 1993, The solar engine and its influence on terrestrial atmosphere and climate (Berlin: Springer).
Proctor, M.R.E, and Gilbert, A.D. (Eds.) 1995, Lectures on Solar and Planetary Dynamos (Cambridge University Press).
Ribes, E. 1990, Phil. Trans. Roy. Soc. London, 330, 487.
Ribes, E. (ed.) 1994, The solar engine and its influence on the terrestrial atmosphere and climate (Springer).
Ribes, E., Merlin, Ph., Ribes, J.-C., & Barthalot, R. 1989, Ann. Geophys. 7. 321.
Rieger, E., Share, G.H., Forrest, D.J., Kanbach, G., Reppin, C., & Chupp, E.L 1984, Nature, 312, 623.
Sakurai, K. 1979, Nature, 278, 146.
Sakurai, K. 1981, Solar Phys., 74, 35.
Scherrer, P.H., et al. 1995, Solar Phys., 162, 129.
Schmitt, D. 1993, The Cosmic Dynamo. Proc. 157th IAU Symp, Potsdam, September 7 - 11, 1992 (eds. Krause, F., Raedler, K.-H., and Ruediger, G.; Dordrecht: Kluwer), 1.

Sokoloff, D., et al. 1993, The solar engine and its influence on terrestrial atmosphere and climate (ed. E. Nesme-Ribes;Berlin: Springer), 99.
Sturrock, P.A., & Walther, G. 1996 (submitted).
Sturrock, P.A., Proc. SOLERS22 Workshop, National Solar Observatory, Sacramento Peak, New Mexico, U.S.A, June 17 - 21, 1996 (to be published in Solar Physics).
Thomson, D.J., Maclennan, C.G., & Lanzerotti, L.J. 1995, Nature, 376, 139.
Toutain, T., & Froelich, C. 1992, Astron. Astrophys. 257. 287.
Walen, C. 1947, Ark. Mat. Atron. Fys., 33A, No. 18.